\documentclass[twocolumn,showpacs,preprintnumbers,amsmath,amssymb]{revtex4}

\usepackage{graphicx}
\usepackage{makecell}
\usepackage{mathrsfs}
\usepackage{bm}

\begin{document}

% the following line is for submission, including submission to the arXiv!!
\hspace{5.2in} \mbox{LA-UR-16-25845}

\title{Neutron-gamma competition for $\beta$-delayed neutron emission}

\author{M. R. Mumpower}
\email{mumpower@lanl.gov}
\affiliation{Theoretical Division, Los Alamos National Laboratory, Los Alamos, NM 87545, USA}

\author{T. Kawano}
% \email{kawano@lanl.gov}
\affiliation{Theoretical Division, Los Alamos National Laboratory, Los Alamos, NM 87545, USA}

\author{P. M\"{o}ller}
% \email{moller@lanl.gov}
\affiliation{Theoretical Division, Los Alamos National Laboratory, Los Alamos, NM 87545, USA}

\date{\today}

\begin{abstract}
We present a coupled Quasi-particle Random Phase Approximation and Hauser-Feshbach (QRPA+HF) model for calculating delayed particle emission. 
This approach uses microscopic nuclear structure information which starts with Gamow-Teller strength distributions in the daughter nucleus, and then follows the statistical decay until the initial available excitation energy is exhausted. 
Explicitly included at each particle emission stage is $\gamma$-ray competition. 
We explore this model in the context of neutron emission of neutron-rich nuclei and find that neutron-gamma competition can lead to both increases and decreases in neutron emission probabilities, depending on the system considered. 
A second consequence of this formalism is a prediction of more neutrons on average being emitted after $\beta$-decay for nuclei near the neutron dripline compared to models that do not consider the statistical decay. 
\end{abstract}

\pacs{23.40.-s, 24.60.Dr}

\maketitle

\section{Introduction}
Delayed neutron emission is a common decay channel found in neutron-rich nuclei.
It was first discovered in 1939 during an investigation of the splitting of uranium and thorium when these elements were bombarded with neutrons \cite{Roberts1939}. 
This phenomenon is energetically possible when a precursor nucleus with $\beta^{-}$-decay energy window ($Q_{\beta}$) populates excited states in the daughter nucleus above the one neutron separation energy ($S_\textrm{n}$). 
The excited states may decay by particle or $\gamma$-ray emission. 
Accurately accounting for $\beta$-delayed neutron emission is necessary in, for example, nuclear reactor operation, to the rapid neutron capture or $r$-process of nucleosynthesis. 

For $r$-process nucleosynthesis, the predicted production pathway ventures far from the currently known nuclides towards the neutron dripline where $S_\textrm{n}$ is small and $Q_{\beta}$ is large.  
In this case, multiple-neutron emission is possible and may even be the dominant decay pathway for many nuclei near the neutron dripline. 
This provides a secondary source of free neutrons during late-times when the nuclei decay back to stability \cite{Mumpower2012a}. 
See Ref. \cite{Mumpower2016-review} for a recent review and further discussions of the impact of $\beta$-delayed neutron emission on $r$-matter flow. 

Experimental studies of $\beta$-delayed neutron emission are difficult due to the challenge of producing short-lived neutron-rich nuclei. 
Studies of these nuclei are further compounded by the requirement of high efficiency neutron detectors \cite{Dillmann2014}. 
Alternative techniques seek to sidestep the problem of developing high efficiency neutron detectors by using recoil-ion spectroscopy coupled with a $\beta$-Paul trap \cite{Scielzo2012}. 
Only recently has it been possible to measure delayed neutron branching ratios for heavy nuclei near the $N=126$ shell closure \cite{Caballero-Folch2016}. 
In total, roughly $\sim250$ measurements exist on neutron emitters in the literature \cite{Dillmann2014}. 

Very few measurements exist for the branching ratios or $P_\textrm{n}$ values of multi-neutron emission. 
The largest known two-neutron emitter ($^{86}$Ga) was observed at the Holifield Radioactive Ion Beam Facility with a $P_{2\textrm{n}}=20(10)$\% \cite{Miernik2013}. 
Future studies at radioactive beam facilities will extend the reach of measureable nuclei well into the regime of multi-neutron emission predicted by theory \cite{Moller+97, Moller+03, Marketin2016}. 
However, it will still be many years before there is sufficient data to fully benchmark theoretical models. 

Delayed neutron emission is challenging to describe theoretically, relying on the complex description of the nucleus, its excited states, and the likelihood of particle emission. 
In particular it is difficult in any model to calculate the energies of the high-lying states near the neutron separation energy very accurately. 
Two approaches to the description of delayed neutron emission can be found in the literature. 
The first approach involves models using systematics, such as the Kratz-Herman formula \cite{KHF1973} or more recent approaches, see e.g. \cite{McCutchan2012}. 
Models using systematics produce a good global fit to known data and improvements have recently been made by Miernik et al. where they provide an approximation to neutron-gamma competition \cite{Miernik2014}. 
The applicability of systematics to multi-neutron emission towards the dripline is unknown, as they are based on simple relations which factor in $Q_{\beta}$ and $S_\textrm{n}$. 

A second approach is to use a microscopic description, usually starting with some form of Quasi-particle Random Phase Approximation (QRPA) that provides the initial population of the daughter nucleus \cite {Moller+97, Moller+03, Borzov+11, Fang+13, Minato+16, Marketin2016}. 
Rapid progress has been made for the description of half-lives using the Finite Amplitude Method which is equivalent to QRPA, but this method has not yet been applied to $P_\textrm{n}$ predictions \cite{Mustonen+16}. 
To calculate $P_\textrm{n}$ values from the QRPA approaches, a schematic formula is used based off the available energy window  \cite{Moller+97, Moller+03, Marketin2016}. 
This approach to calculating $P_\textrm{n}$ values is sometimes referred to the `cutoff' method in the literature. 
For ease of citation later in the paper, we define three reference models: QRPA-1 (Gamow-Teller (GT) only) \cite{Moller+97} and QRPA-2 (additional inclusion of First-Forbidden (FF) transitions) \cite{Moller+03}, and D3$C^{*}$ \cite{Marketin2016}. 
The latter is a microscopic approach based off covariant density functional theory and proton-neutron relativistic QRPA. 

In this paper we present a further upgrade to the microscopic framework of computing delayed neutron-emission probabilities. 
Our approach combines both the QRPA and statistical methods into a single theoretical framework. 
This is achieved by first calculating the $\beta$-decay intensities to accessible states in the daughter nucleus using QRPA. 
The subsequent decay of these states by neutron or $\gamma$-ray emission is then treated in the statistical Hauser-Feshbach (HF) theory. 
This work extends that of Ref. \cite{Kawano+08}, which was limited to near stable isotopes, to the entire chart of nuclides. 

\section{Model}
\begin{figure*}
 \begin{center}
  \centerline{\includegraphics[width=150mm]{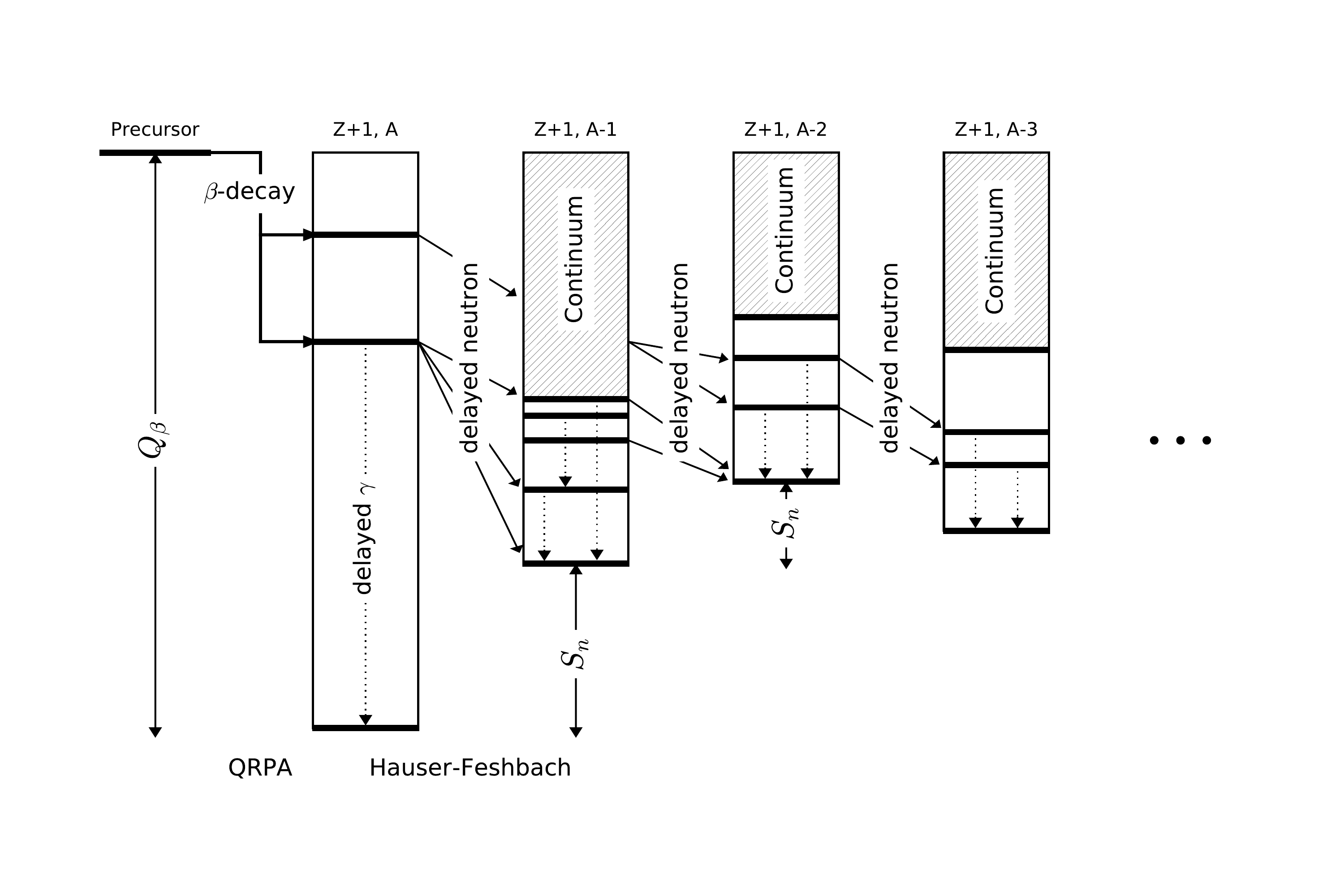}}
  \caption{\label{fig:model} Schematic of the combined QRPA+HF approach. Initial population of the the daughter nucleus ($Z+1$,$A$) is determined by QRPA. Subsequent delayed-neutron and $\gamma$-ray emission are handled in the HF framework and shown by solid and dotted lines respectively. The statistical decay is followed until all available excitation energy is exhausted, denoted by trailing dots. }
 \end{center}
\end{figure*}

The calculation of $\beta$-delayed neutron emission in the Quasi-particle Random Phase Approximation plus Hauser-Feshbach (QRPA+HF) approach is a two step process.
It involves both the $\beta$-decay of the precursor nucleus ($Z$,$A$) where $Z$ represents the atomic number, and $A$ is the total number of nucleons, and the statistical decay of the subsequent generations, ($Z+1$,$A-j$) where $j$ is the number of neutrons emitted, ranging from $0$ to many. 
The statistical decay continues until all the available excitation energy is exhausted. 
This is shown schematically in Fig \ref{fig:model}. 

Beginning with the precursor nucleus ($Z$,$A$), the decay to the daughter ($Z+1$, $A$) is calculated using the QRPA framework of M\"{o}ller et al. \cite{Krumlinde+84, Moller+90, Moller+97, Moller+03}. 
In this framework, the Schr\"{o}dinger equation is solved for the nuclear wave functions and single-particle energies using a folded-Yukawa Hamiltonian. 
Residual interactions (pairing and Gamow-Teller) are added to obtain the decay matrix element from initial to final state, $\langle f|\beta_{GT}| i\rangle$. 
A $\beta$-decay strength function is used to describe the behavior of the squares of overlap integrals of nuclear wave functions. 
The $\beta$-decay strength function is given by \cite{Hansen+73}
\begin{equation}
   S_\beta(E_x) = \frac{I(E)}{f(Z,Q_{\beta}-E_x)T_{1/2}}
   \label{eqn:S_beta}
\end{equation}
where $I(E_x)$ $\beta$-intensity per MeV of levels at energy $E_x$ in the daughter nucleus, $f$ is the usual Fermi function and $T_{1/2}$ is the half-life of the parent nucleus in seconds. 
The units of $S_\beta$ are $s^{-1}MeV^{-1}$. 
Here we only consider the Gamow-Teller (GT) strength from QRPA, similar to Ref. \cite{Moller+97}. 

The transition energies and rates for $\beta$-decay strongly depend on the level structure in the participating nuclei. 
Due to the natural uncertainty associated with this calculation we smooth the $\beta$-strength distribution by a Gaussian,
\begin{equation}
  \omega(E_x) = C \sum_k b^{(k)}
     {1 \over{\sqrt{2\pi} \Gamma}}
     \exp\left\{
            - {{ [E^{(k)} - E_x]^2 }\over{2\Gamma^2}}
         \right\},
  \label{eq:omega}
\end{equation}
where $b^{(k)}$ are the branching ratios from the parent state to the daughter states $E^{(k)}$, $E_x$ is the excitation energy of the daughter nucleus, and the Gaussian width is taken to $\Gamma=100$ keV, as in \cite{Kawano+08}. 
The normalization constant, $C$, is the given by the condition
\begin{equation}
 \int_{0} ^{Q_\beta}\omega(E_x) \textrm{d}E_x = 1 ,
  \label{eq:omeganorm}
\end{equation}
where $Q_{\beta}$ is the $\beta^{-}$-decay energy window. 

Our QRPA solution may be fully replaced or supplemented with experimental data when available. 
In the case where the level structure and $\beta$-decay scheme are known completely, data are taken from the RIPL-3 / ENSDF and no QRPA strength is used. 
If the database is incomplete, we combine the QRPA solutions with renormalized ENSDF data. 
In some cases, energy levels are known but the spin and parity assignments are uncertain. 
In this case, or in the case of that no level information is found in ENSDF, the QRPA solution is relied upon. 

When applicable, ground state masses are taken from the latest compilation of the Atomic Mass Evaluation (AME2012) \cite{Audi+12}. 
Ground state masses and deformations from the 1995 version of the Finite-Range Droplet Model (FRDM1995) are used unless otherwise noted \cite{FRDM1995}. 

In the second stage of the calculation we assume the daughter nucleus is a compound nucleus initially populated by the $\beta$-decay strength from the QRPA calculations. 
This assumption is based off the independence hypothesis from N. Bohr, from which it follows that the compound nucleus depends only on its overall properties rather than the details of formation \cite{Weisskopf+56}. 
We follow the statistical decay of this nucleus and subsequent generations until all the available excitation energy is exhausted using the Los Alamos CoH statistical Hauser-Feshbach code \cite{Kawano+08, Kawano+10, Kawano+16}. 

For calculation of the neutron transmission coefficients we employ the Koning-Delaroche global optical potential that has an iso-spin dependent term \cite{Koning+03}. 
This model has been shown to provide a very good description of neutron-rich nuclei, see e.g. Ref. \cite{Koning+05, Goriely+08}. 
We have also used the Kunieda deformed optical potential \cite{Kunieda+07}, however we find only minor changes to our results and so do not report on this model here. 

As shown in Fig.~\ref{fig:model}, the excited states in the subsequent generations can be discrete or in the continuum. 
Above the last known level, or in the case of no known levels, the Gilbert-Cameron level density is used \cite{Gilbert1965}. 
This hybrid formulation uses a constant temperature model at low energies and matches to a Fermi gas model in the high energy regime. 
Parameters for this formula are taken from systematics at stability and extrapolated as one crosses the nuclear landscape \cite{Kawano2006}.
Shell corrections are included in the level density via the method proposed by Ignatyuk et al. \cite{Ignatyuk1975}. 

The $\gamma$-ray transmission coefficients are constructed using the generalized Lorentzian $\gamma$-strength function ($\gamma$SF) for E1 transitions. 
Higher order electric and magnetic transitions are also considered, but due to the strong dependence of the transitions on multipolarity, lower $L$ transitions usually take precedence. 

A key component of the HF calculations is the competition of $\gamma$-ray emission at each stage of particle emission. 
We now introduce several definitions to make this discussion more precise. 
We denote the compound state as $c^{(j)}_k$ where $c^{(j)}$ represents the $j$-th compound nucleus and $k$ the index of the excited state in this nucleus which takes values from the highest excitation to the ground state. 
Using this definition, $c^{(0)}$ would be shorthand for describing the daughter nucleus ($Z+1$,$A$) and $c^{(j)}$ would describe ($Z+1$,$A-j$). 

The $\gamma$-emission transition probability in the $j$-th compound nucleus is defined as
\begin{equation}
 p_j(E_i, E_k) = \frac{1}{N_j(E_i)} T^{(j)}_\gamma(E_i-E_k) \rho_j(E_k)
 \label{eqn:pprob}
\end{equation}
where the transition is from a level with high excitation energy $E_i$ to a level of energy $E_k$, $T^{(j)}_\gamma$ is the $\gamma$-ray transmission coefficient for $c^{(j)}$, $\rho_j(E_k)$ is the level density in $c^{(j)}$ at energy $E_k$ and $N_j(E_i)$ is a normalization factor defined below. 
The neutron-emission transition probability from $c^{(j)}$ to $c^{(j+1)}$ is defined as
\begin{equation}
 q_j(E_i, E_{k^\prime}) = \frac{1}{N_j(E_i)} T^{(j+1)}_n(E_i-S^{(j)}_n-E_{k^\prime}) \rho_{j+1}(E_{k^\prime})
 \label{eqn:qprob}
\end{equation}
where the transition is from an energy level $E_i$ in $c^{(j)}$ to an energy level $E_k$ in  $c^{(j+1)}$, $S^{(j)}_n$ is the neutron separation energy of $c^{(j)}$, $T^{(j+1)}_n$ is the neutron transmission coefficient from $c^{(j+1)}$ to $c^{(j)}$ and $\rho_{j+1}(E_{k^\prime})$ is the level density in $c^{(j+1)}$ evaluated at $E_{k^\prime}$. 
The prime on the $k$ index indicates that the energy level is in a different compound nucleus. 
The normalization factor, $N_j$ is given by the sum over all possible exit channels
\begin{multline}
 N_j(E_i) = \int_0^{E_i} T^{(j)}_{\gamma}(E_i-E_k) \rho_j(E_k) \textrm{d}E_k \\
               + \int_0^{{E_i}-S^{(j)}_n} T^{(j+1)}_{n}(E_i-S^{(j)}_n-E_{k^\prime}) \rho_{j+1}(E_{k^\prime}) \textrm{d}E_{k^\prime}
 \label{eqn:normprob}
\end{multline}
where the integration in each case runs over the appropriate energy window. 
In the above equations we have explicitly omitted the indices of quantum numbers and assumed implicitly that all transitions obey spin-parity selection rules. 
The units for $q$ and $p$ are in [1/MeV]. 
Also note that in both Eq. (\ref{eqn:pprob}) and (\ref{eqn:qprob}) the transition probabilities do not depend on how the initial state was populated in accordance with the compound nucleus assumption. 

The level population in $c^{(j+1)}$ for particular energy $E_k$ is defined as 
\begin{eqnarray}
 \mathscr{P}_{j+1}(E_k) = \sum_{i=0}^{k-1} \mathscr{P}_{j+1}(E_i) p_{j+1}(E_i, E_k)\nonumber \\ 
                        + \sum_{{k^\prime}=0}^{k-1} \mathscr{P}_j(E_{k^\prime}) q_j(E_{k^\prime}, E_k)
 \label{eqn:lvlpop}
\end{eqnarray}
where the summations run over all levels which may feed the compound state $c^{(j+1)}_k$. 
The first term takes into account $\gamma$ emission in $c^{(j+1)}_k$ while the second term takes into account neutron emission from $c^{(j)}_{k^\prime}$. 
Both terms depend on all the possible pathways taken to populate $E_k$. 
Thus, the population of a particular level is a recursive function which can be handled naturally on a computer. 
In our case, the initial population comes from $\beta$-decay, such that $\mathscr{P}_0(E_k)=S_\beta(E_k)$. 

From these definitions, we can compute the $j$-th neutron emission probability by considering all the channels that end in the $j$-th compound nucleus relative to all possible pathways:
\begin{equation}
 P_{j\textrm{n}} = \mathscr{P}_j(E_{\textrm{gs}})
 \label{eqn:Pjn}
\end{equation}
where we evaluate the level population at the ground state energy, $E_{\textrm{gs}}$, in $c^{(j)}$. 
This equation is a recursive convolution and therefore in general a $P_{j\textrm{n}}$ value may increase or decrease relative to calculations based on the older energy window argument. 
Beta-decay with no neutron emission is denoted as $P_{0\textrm{n}}$ and because of the normalization in the calculation of the level population we have a similar constraint on the neutron emission probabilities, $\sum_{j=0} P_{j\textrm{n}} = 1$. 

Delayed neutron spectra can also be obtained from the coupled QRPA+HF approach. 
This has already been explored in the context of the QRPA+HF model in a previous study \cite{Kawano+08} so we do not cover this here. 

\section{Results}
We have performed QRPA+HF calculations to obtain new delayed-neutron emission probabilities for all neutron-rich nuclei from the valley of stability towards the neutron dripline. 
Our aim in this section is to present a general overview of the model results. 
We also discuss quantitative benchmarks of the model performance, but we caution that most measured systems do not exhibit large neutron-gamma competition and so we do not expect a large deviation of our results from previous models \cite{Moller+97, Moller+03}. 
However, far from the stable isotopes we find substantial deviations from older model predictions and discuss the significance of this below. 

\subsection{Neutron-gamma competition}

\begin{figure}
 \begin{center}
  \centerline{\includegraphics[width=85mm]{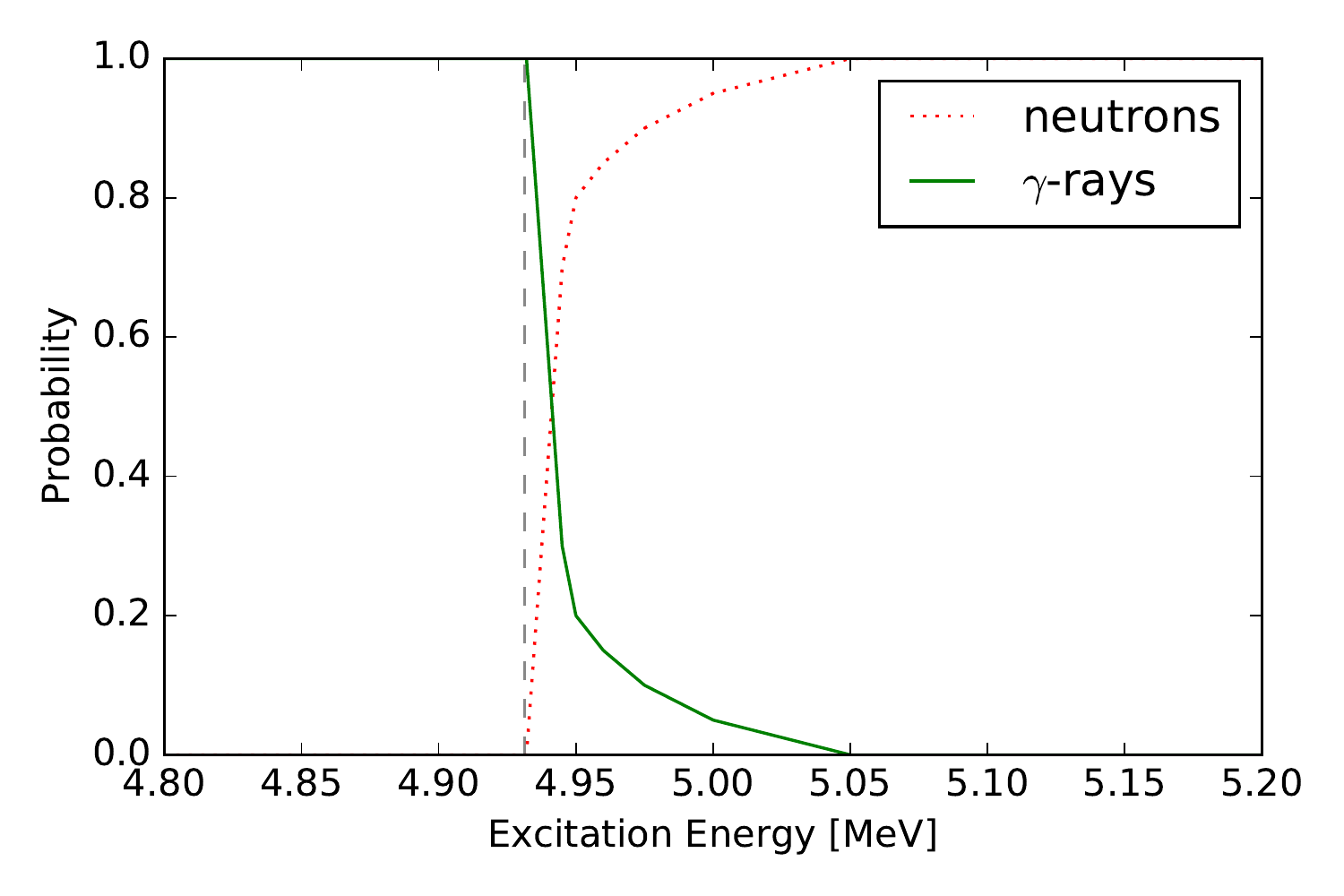}}
  \caption{\label{fig:ngcomp} (Color online) Neutron-gamma competition in $^{86}$Ge after $\beta$-decay of $^{86}$Ga. The dotted line represents the neutron separation energy of $^{86}$Ge. }
 \end{center}
\end{figure}

Figure \ref{fig:ngcomp} shows a representative example of neutron-gamma competition for $^{86}$Ge ($Z=32$) which is populated after the $\beta$-decay of $^{86}$Ga ($Z=31$). 
From the transition probabilities near the one neutron separation energy in $^{86}$Ge we see that the neutron-gamma competition extends roughly $100$ keV above $S_\textrm{n}=4.93$. 

\begin{figure}
 \begin{center}
  \centerline{\includegraphics[width=85mm]{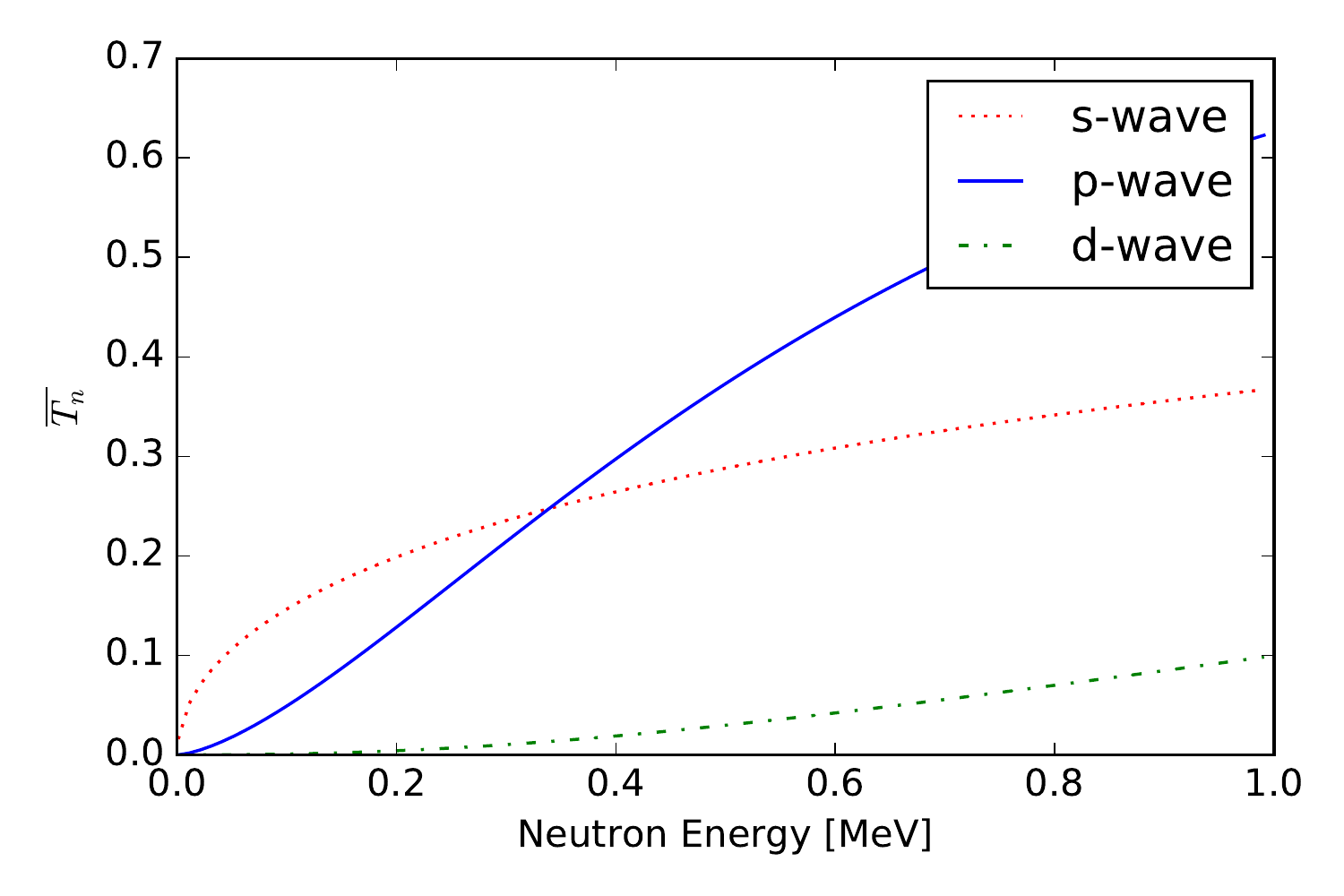}}
  \caption{\label{fig:ntrans} (Color online) Neutron transmission coefficients for s, p and d-wave neutron emitted from $^{86}$Ge. For p and d waves the transmission for different spins are averaged. }
 \end{center}
\end{figure}

The neutron transmission coefficients for s, p and d-waves are shown in Fig.~\ref{fig:ntrans}. 
For p and d wave neutrons the transmissions for different spins are averaged. 
A quick rise in p-wave neutrons produces the sharp cut-off of neutron-gamma competition observed in Fig.~\ref{fig:ngcomp}. 
Despite this small energy range for neutron-gamma competition, there is still an impact on the predicted $P_\textrm{n}$ values. 
The QRPA-1 model predicts $P_{1\textrm{n}}\sim61$\% and $P_{2\textrm{n}}\sim13$\% while QRPA+HF yields $P_{1\textrm{n}}\sim69$\% and $P_{2\textrm{n}}\sim6$\%. 
While there is a larger discrepancy with experiment, $P_{2\textrm{n}}\sim20(10)$\%, this result is still within the mean model error, which we will discuss below. 
We further note that it is likely that FF transitions play a role in this decay, which have not been considered in either calculation. 
When FF transitions are considered, using the QRPA-2 model, there is a shift to larger neutron emission for $^{86}$Ga: $P_{1\textrm{n}}\sim20$\% and $P_{2\textrm{n}}\sim44$\%
This example highlights the delicate dependency of $P_\textrm{n}$ predictions on the $\beta$-strength function. 

\begin{figure}
 \begin{center}
  \centerline{\includegraphics[width=95mm]{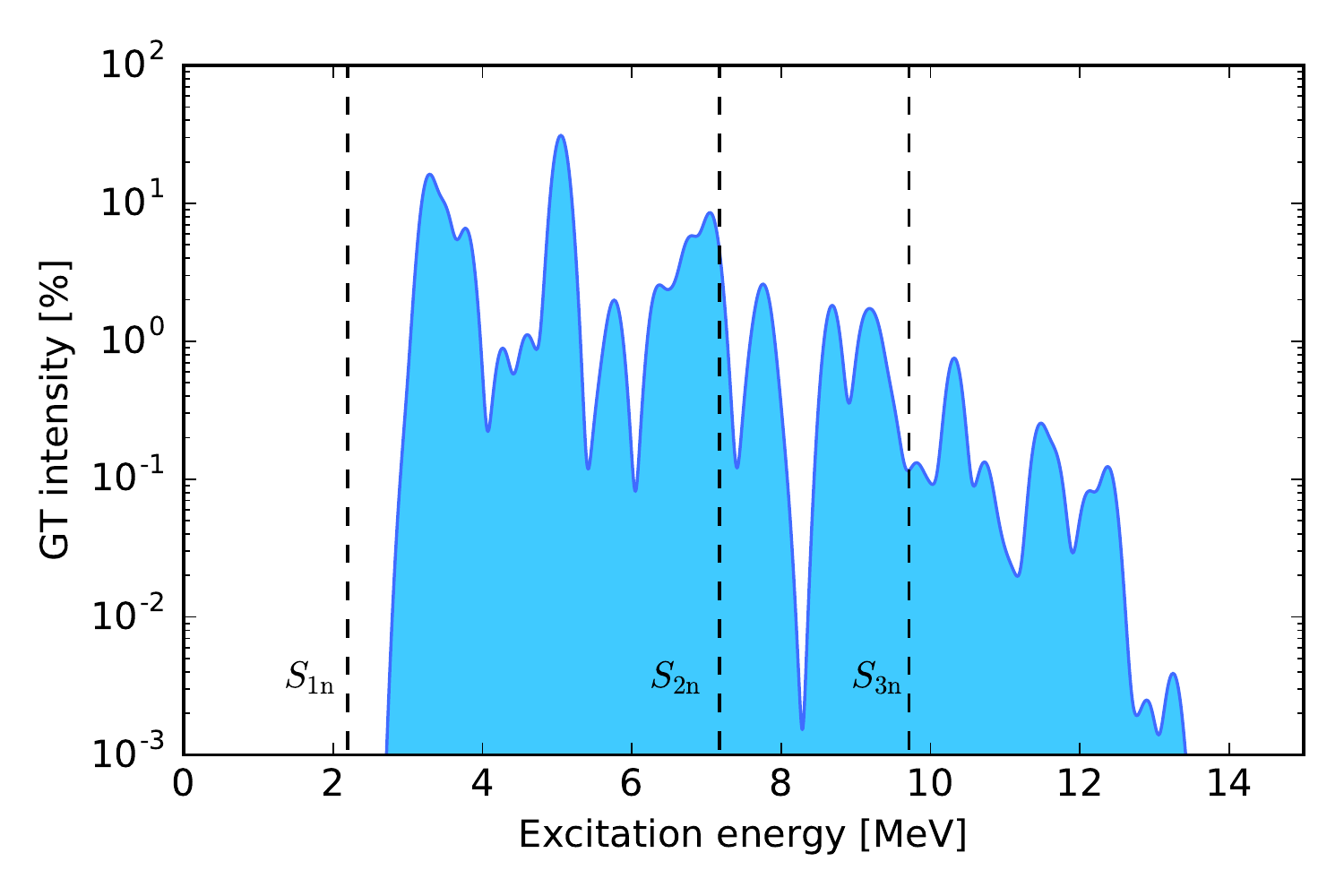}}
  \caption{\label{fig:As93} (Color online) Gaussian broadened Gamow-Teller $\beta$-decay intensity of $^{93}$As from the QRPA-1 model of Ref. \cite{Moller+97}. }
 \end{center}
\end{figure}

It is difficult to predict whether or not $P_\textrm{n}$ values will increase or decrease relative to the QRPA-1 model due to the convolution involved in Eq. (\ref{eqn:Pjn}). 
Consider the case of $^{93}$As. 
The Gaussian broadened GT $\beta$-decay intensity is shown in Fig.~\ref{fig:As93}. 
From the energy window argument, one obtains $P_{1\textrm{n}}\sim84$\%, $P_{2\textrm{n}}\sim14$\% and $P_{3\textrm{n}}\sim2$\% with the QRPA-1 model. 
The new QRPA+HF model predicts an overall shift to smaller neutron emission: $P_{1\textrm{n}}\sim94$\%, $P_{2\textrm{n}}\sim6$\% and $P_{3\textrm{n}}\sim0$\%, which is similar to the previous case of the $\beta$-decay of $^{86}$Ga. 

The reason for this reduction typically comes from the suppression of neutron emission just above threshold due to spin-parity conservation. 
For the vast majority of nuclei near stability, roughly $75$\%, we find a reduction of $P_{1\textrm{n}}$ relative to the QRPA-1 model. 
In contrast, we find that some nuclei, roughly $25$\%, have their neutron emission increased. 
An example is $^{145}$Cs where the QRPA-1 model predicts $P_{0\textrm{n}}\sim98$\% and $P_{1\textrm{n}}\sim2$\%. 
The QRPA+HF model predicts $P_{0\textrm{n}}\sim78$\% and $P_{1\textrm{n}}\sim22$\%, which is closer to the measured value of $P^{exp}_{1\textrm{n}}\sim15$\%. 

Most recently it has been shown that there is significant neutron-gamma competition above the neutron threshold associated with the $\beta$-decay of $^{70}$Co ($Z=27$) using the Total Absorption Gamma Spectroscopy (TAGS) technique \cite{Spyrou+16}. 
We do find neutron-gamma competition in the daughter nucleus $^{70}$Ni ($Z=28$) in the QRPA+HF model, however, our results are dependent on the uncertain spin-parity assignments in ENSDF, which makes it difficult to draw any concrete conclusions. 
If the Gilbert-Cameron level density is used for this calculation, neutron emission will dominant as there is always an open pathway for neutron emission to proceed at high excitation energies. 
This nucleus highlights the importance of using realistic energy levels as well as spin-parity assignments in the calculation of delayed neutron emission probabilities. 

\begin{figure}
 \begin{center}
  \centerline{\includegraphics[width=95mm]{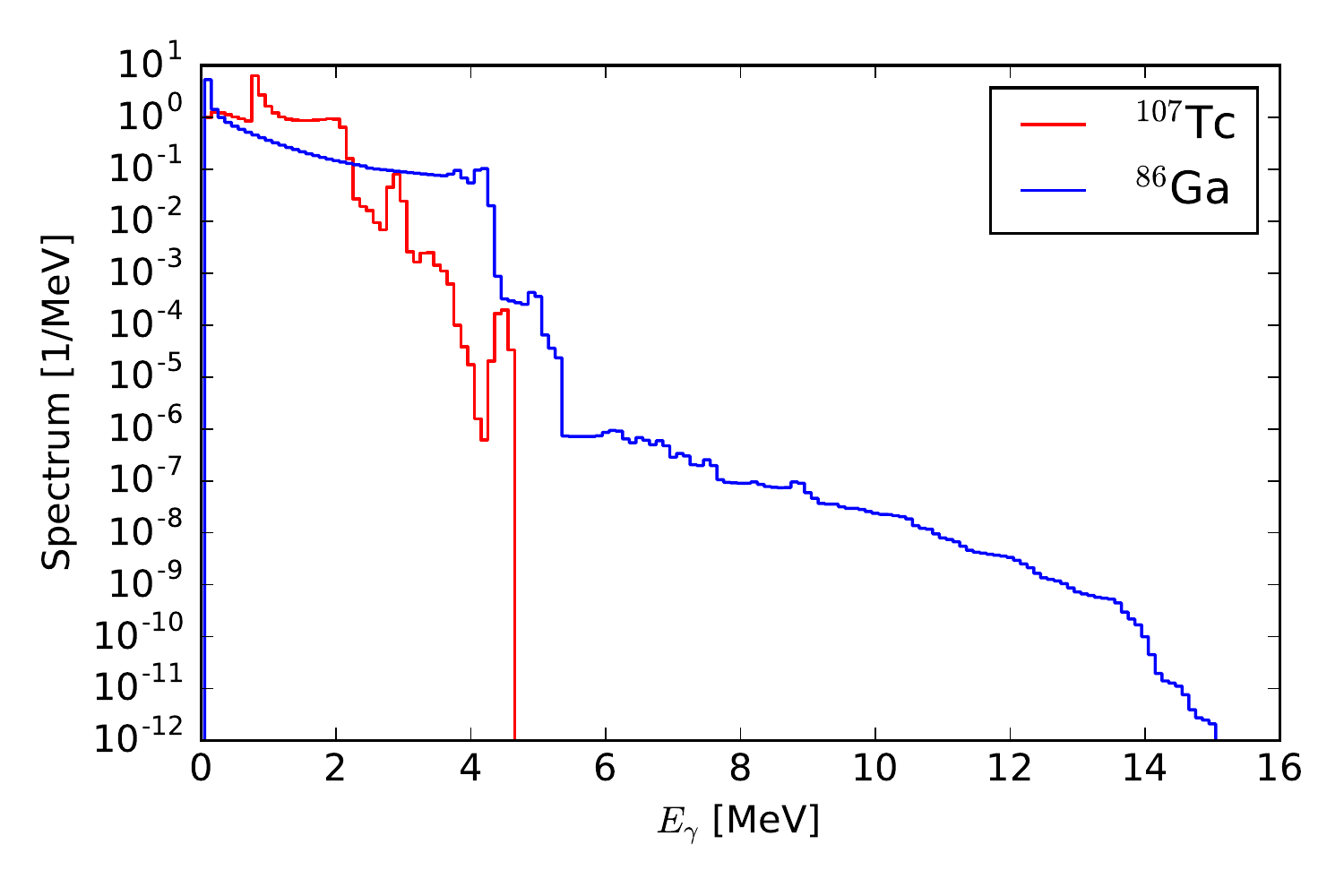}}
  \caption{\label{fig:gammas} (Color online) The $\gamma$-ray spectrum after the $\beta$-decay of $^{107}$Tc and $^{86}$Ga. }
 \end{center}
\end{figure}

The TAGS technique coupled with the production potential of radioactive beam facilities will allow for the exploration of nuclear structure through $\beta$-decay measurements far from stability. 
To this end the QRPA+HF framework can help to isolate interesting nuclei by constructing the expected $\gamma$-spectrum associated with the decay. 
We show the capacity of the QRPA+HF in predicting $\gamma$-spectrum in Fig.~\ref{fig:gammas} for select nuclei: $^{107}$Tc and $^{86}$Ga. 
A clear distinction can be seen between these two cases. 
The $\beta$-decay of $^{107}$Tc produces no neutron emission resulting in the distinct drop off of the $\gamma$-spectrum around $4.65$ MeV whereas $^{86}$Ga is our familiar two neutron emitter discussed above. 
The long tail at high energies in the $\gamma$-spectrum of this decay comes from the competition of neutron and $\gamma$ emission while the sharp drops are an indication that neutron emission has occurred. 

\subsection{Global model results}

\begin{figure*}
 \begin{center}
  \centerline{\includegraphics[width=175mm]{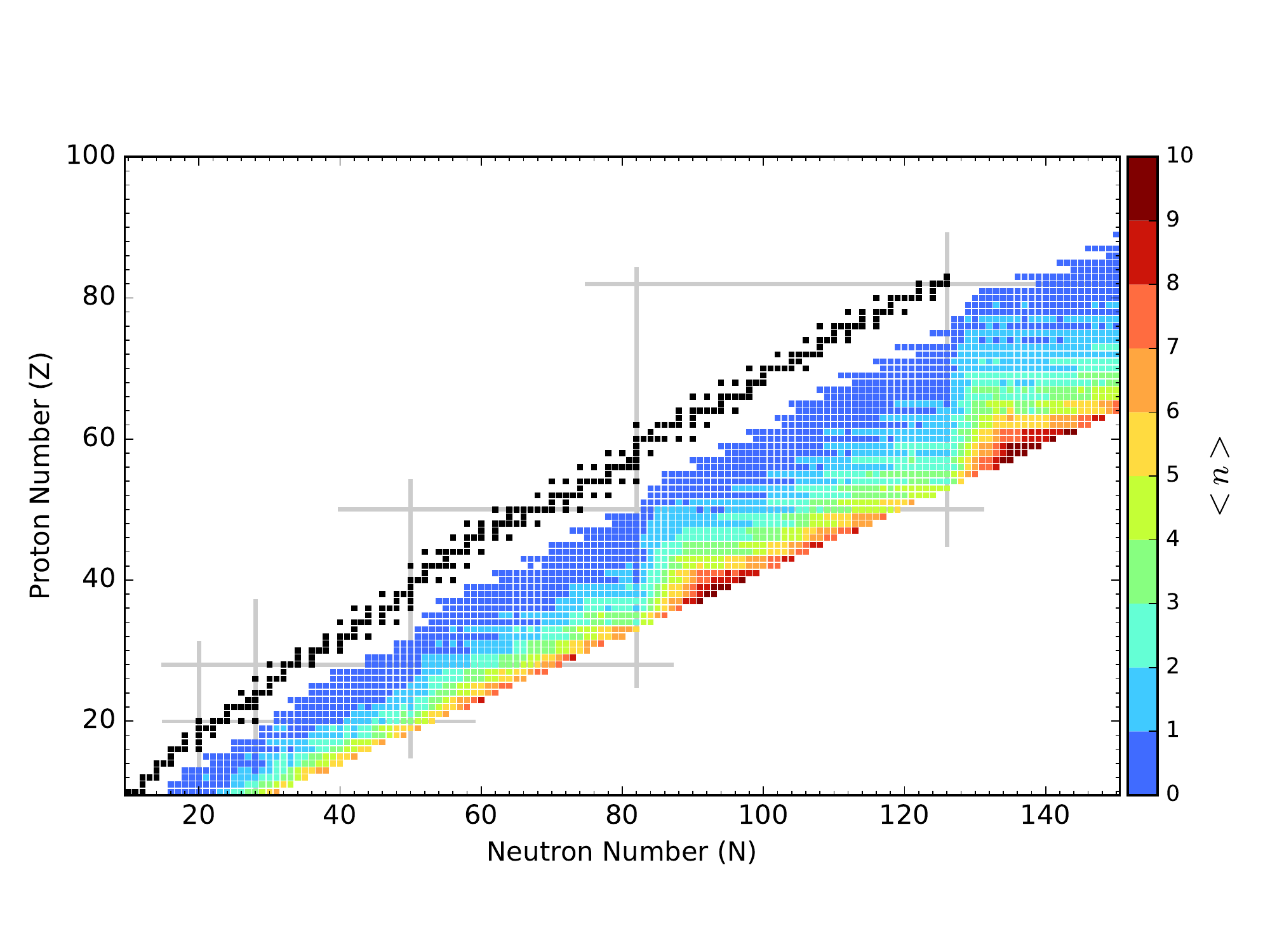}}
  \caption{\label{fig:predictions} (Color online) Average number of neutrons emitted after a single $\beta$-decay using the QRPA+HF approach with FRDM1995 masses and GT strength data from Ref. \cite{Moller+97}. Closed shells shown by parallel straight lines with stable nuclei shown in black. }
 \end{center}
\end{figure*}

To show our results globally across the chart of nuclides in a concise manner we compute the average number of delayed neutrons per $\beta$-decay 
\begin{equation}
 \langle n \rangle = \sum_{j=0} j P_{j\textrm{n}}
\end{equation}
where $j$ denotes the number of neutrons emitted, $P_{j\textrm{n}}$ is the computed probability to emit $j$-neutrons as defined in Eq. (\ref{eqn:Pjn}) and the sum of $P_{j\textrm{n}}$ is unity. 
The delayed-neutron emission predictions using the QRPA+HF framework throughout the chart of nuclides are shown in Fig.~\ref{fig:predictions}. 
We find that neutron-emission channel begins on average roughly five or so units in neutron number from the last stable isotope, in agreement with measurements. 
Figure \ref{fig:predictions} shows a fairly smooth trend in the number of delayed neutrons as neutron excess increases, which mirrors the results of older model calculations \cite{Moller+97, Moller+03}. 

For the majority of nuclei near stability, we find neutron emission to dominate over $\gamma$ emission above the neutron threshold. 
For these nuclei, it is okay to assume that decays to energies above the separation energy always lead to delayed neutron emission. 
To date, this has been the basis for model calculations of delayed-neutron emission probabilities, see e.g. Eq. (2) in Ref. \cite{Moller+03}. 
Towards the dripline however, $j$-neutron separation energies, $S_{j\textrm{n}}$, may overlap causing the assumption to become difficult to implement as it is not clear how to partition the $\beta$-decay strength without double counting. 
The older models, QRPA-1 and QRPA-2, handled this by mapping any additional neutron emission above $j$-neutrons into the $j$-neutron emission probability, where $j=3$.  

We illustrate this scenario by the transition between ($Z+1$,$A-2$) and ($Z+1$,$A-3$) in Fig.~\ref{fig:model}. 
Here, the energy window argument becomes invalid and only the QRPA+HF model provides the correct treatment via the statistical decay. 
Thus, proper treatment of delayed neutron emission probabilities requires neutron-gamma competition as one approaches the neutron dripline. 
This can be seen most notably after magic neutron numbers, indicated by solid vertical lines in Fig.~\ref{fig:predictions}. 
We find an enhancement in the number of neutrons emitted in these regions compared to the aforementioned method.
This is where the largest deviation from older models occurs, with neutron emission exceeding $\langle n \rangle \sim 5$ or more in QRPA+HF. 
Whether this has an impact on $r$-matter flow will depend on the shape of the neutron dripline among other things. 
Complete reaction network calculations must be used in order to determine the full impact on the $r$-process and we reserve this effort for a future study. 

\subsection{Model comparisons}

We now discuss our results in a quantitative fashion. 
Since $P_\textrm{n}$ values range from $0$\% to $100$\%, it is easiest to consider model residuals or differences between theoretical calculations and experimental data. 
Other physical quantities, for example half-lives, vary orders of magnitude. 
Further, the \textit{differences} between calculation and experiment can also vary orders of magnitude. 
Ergo, other means of analysis must be considered when dealing with such quantities \cite{Moller+03}. 

To gauge the average residuals of a given model we compute
\begin{equation}
 \bar{e} = \frac{1}{N} \sum^{N}_{i=1} | P_{1\textrm{n}}(i) - P^{exp}_{1\textrm{n}}(i) |
\end{equation}
where the sum runs over $N=225$ measured $P_{1\textrm{n}}$ values. 
The QRPA-1 model of Ref. \cite{Moller+97} yields an average deviation $\bar{e}_{QRPA-1}=15.4$\% and for our new QRPA+HF we obtain $\bar{e}_{QRPA+HF}=15.3$\%. 
As expected, the inclusion of neutron-gamma competition does not greatly impact the results near stability. 
The latest fully microscopic approach from Ref. \cite{Marketin2016} gives comparable predictability, $\bar{e}_{D3C^{*}}=15.4$\%. 
This microscopic model also includes FF transitions. 
If FF transitions are considered in the macroscopic-microscopic approach, as in the QRPA-2 model of Ref. \cite{Moller+03}, a substantial improvement is seen in the mean error $\bar{e}_{QRPA-2}=12.2$\%.
This suggests that FF transitions play an important factor in improving the description of $P_{j\textrm{n}}$ values. 
An update to the QRPA-2 model which includes the neutron-gamma competition discussed here will be the subject of an upcoming publication. 

\bgroup
\begin{table}[]
\centering
\caption{Number of $P_{1\textrm{n}}$ theory predictions within $\pm x$\% of measurements for various models. The label QRPA+HF represents the results of this work. }
\label{tab1}
\def\arraystretch{1.5}      % Stretch table cell row whitespace padding
\setlength\tabcolsep{1.5ex} % Stretch table cell column whitespace padding
\begin{tabular}{l|c|c|c|c|c|c}
             & 5\% & 10\% & 20\% & 30\% & 40\% & $>$40\%  \\ \Xhline{1.5pt}
QRPA+HF      & 115 & 21   & 29   & 21   & 10   & 29       \\ \hline
QRPA-1       & 113 & 23   & 34   & 16   & 11   & 28       \\ \hline
QRPA-2       & 117 & 24   & 35   & 20   & 9    & 20       \\ \hline
D3C$^{*}$    & 98  & 28   & 34   & 26   & 12   & 27              
\end{tabular}
\end{table}
\egroup

It is also instructive to consider how many $P_{1\textrm{n}}$ theory predictions are within $\pm x$\% of measurements. 
This metric is summarized for several models in Table \ref{tab1}. 
An inspection of the table reveals that roughly $50$\% of $P_{1\textrm{n}}$ predictions are within $\pm 5$\% of measurements. 
The remainder of $P_{1\textrm{n}}$ predictions fall in the tail of the distribution, greater than $\pm 10$\%, which contribute significantly to the mean model error. 
The cause of strong outliers greater than $40$\% is most likely due to a poor reproduction of the $\beta$-strength. 

There are many contributing factors that may impact the calculation of $P_\textrm{n}$ values. 
We now consider the sensitivity of our results to the $\gamma$SF. 
An increase to the $\gamma$SF of roughly one order of magnitude has been recently suggested from emission of $\gamma$-ray in $^{87,88}$Br and $^{94}$Rb. by J. Tain et al. \cite{Tain+15}. 
We have performed the suggested increase of one order of magnitude to the $\gamma$SF used in QRPA+HF and find our results are mostly unchanged. 
The largest change is roughly $\Delta P_{1\textrm{n}}=3$\%, with nearly all nuclei showing little to no change in $P_\textrm{n}$ values. 
This is quite predictable outcome, as $\gamma$-ray transmission is generally orders of magnitude weaker than neutron transmission. 
We conclude from the perspective of delayed neutron emission that an enhancement of an order of magnitude is not warranted for the $\gamma$SF. 
We recognize that such an enhancement would have a larger effect on the neutron capture rates of neutron-rich nuclei. 
This is because cross sections are proportional to $\frac{T_n T_{\gamma}}{T_n + T_{\gamma}}$ and $T_n\sim T_n + T_{\gamma}$. 
Thus, a relatively small change to $T_\gamma$ will have a considerable impact on the cross section. 

\section{Summary}
We have presented a more microscopic method to calculate delayed particle emission. 
This approach couples the QRPA and HF formalisms into one theoretical framework (QRPA+HF) which permits the calculation of $\gamma$-spectra as well as delayed particle spectra and probabilities. 

We explored this new framework in the context of delayed neutron emission and find only a slight improvement in neutron emission probabilities near stability. 
Towards the neutron dripline we find substantial differences from older model predictions due to the competition of neutrons and $\gamma$-rays above the neutron threshold, which will likely have an influence on $r$-matter flow. 
We found the prediction of $P_\textrm{n}$ values depends strongly on the $\beta$-strength function and thus on the associated model used to construct the initial population in the daughter nucleus. 
The $\gamma$SF plays a lesser role, only impacting $P_\textrm{n}$ values up to $\sim3$\% and we observed a negligible impact from the choice of the HF neutron optical model. 

A proper theoretical description of $P_\textrm{n}$ values depends on many facets of nuclear physics from nuclear structure to the reaction mechanism. 
Future $P_\textrm{n}$ measurements will provide great insight into inner workings of the atomic nucleus and with sufficient data lead to the ability to properly benchmark theoretical models. 
Prospective studies are planned to explore the impact the QRPA+HF model has on nucleosynthesis calculations. 

\section{Acknowledgements}
We thank Iris Dillmann for helpful discussions and her student, Stephine Ciccone, for providing a database of recent $P_\textrm{n}$ measurements. 
We thank Scott Marley for helpful discussions regarding current experimental techniques. 
This work was carried out under the auspices of the National Nuclear Security Administration of the U.S. Department of Energy at Los Alamos National Laboratory under Contract No. DE-AC52-06NA25396. 

\bibliographystyle{unsrt}
\bibliography{refs}

\end{document}